\documentclass[twoside,twocolumn,9pt]{article}
\usepackage{extsizes}
\usepackage[super,sort&compress,comma]{natbib} 
\usepackage[version=3]{mhchem}
\usepackage[left=1.5cm, right=1.5cm, top=1.785cm, bottom=2.0cm]{geometry}
\usepackage{balance}
\usepackage{mathptmx}
\usepackage{sectsty}
\usepackage{graphicx} 
\usepackage{amssymb}
\usepackage{amsmath}
\usepackage{amsfonts}
\usepackage{lastpage}
\usepackage[format=plain,justification=justified,singlelinecheck=false,font={stretch=1.125,small,sf},labelfont=bf,labelsep=space]{caption}
\usepackage{float}
\usepackage{fancyhdr}
\usepackage{fnpos}
\usepackage[english]{babel}
\addto{\captionsenglish}{%
  
}
\usepackage{array}
\usepackage{droidsans}
\usepackage{charter}
\usepackage[T1]{fontenc}
\usepackage[usenames,dvipsnames]{xcolor}
\usepackage{setspace}
\usepackage[compact]{titlesec}
\usepackage{hyperref}
\usepackage{epstopdf}%This line makes .eps figures into .pdf - please comment out if not required.

\NewCommandCopy{\amshbar}{\hbar}

\definecolor{cream}{RGB}{222,217,201}

\begin{document}

\pagestyle{fancy}
\thispagestyle{plain}
\fancypagestyle{plain}{
%%%HEADER%%%
\renewcommand{\headrulewidth}{0pt}
}
%%%END OF HEADER%%%

%%%PAGE SETUP - Please do not change any commands within this section%%%
\makeFNbottom
\makeatletter
\renewcommand\LARGE{\@setfontsize\LARGE{15pt}{17}}
\renewcommand\Large{\@setfontsize\Large{12pt}{14}}
\renewcommand\large{\@setfontsize\large{10pt}{12}}
\renewcommand\footnotesize{\@setfontsize\footnotesize{7pt}{10}}
\makeatother

\renewcommand{\thefootnote}{\fnsymbol{footnote}}
\renewcommand\footnoterule{\vspace*{1pt}% 
\color{cream}\hrule width 3.5in height 0.4pt \color{black}\vspace*{5pt}} 
\setcounter{secnumdepth}{5}

\makeatletter 
\renewcommand\@biblabel[1]{#1}            
\renewcommand\@makefntext[1]% 
{\noindent\makebox[0pt][r]{\@thefnmark\,}#1}
\makeatother 
\renewcommand{\figurename}{\small{Fig.}~}
\sectionfont{\sffamily\Large}
\subsectionfont{\normalsize}
\subsubsectionfont{\bf}
\setstretch{1.125} %In particular, please do not alter this line.
\setlength{\skip\footins}{0.8cm}
\setlength{\footnotesep}{0.25cm}
\setlength{\jot}{10pt}
\titlespacing*{\section}{0pt}{4pt}{4pt}
\titlespacing*{\subsection}{0pt}{15pt}{1pt}
%%%END OF PAGE SETUP%%%

%%%FOOTER%%%
\fancyfoot{}
\fancyfoot[LO,RE]{\vspace{-7.1pt}\includegraphics[height=9pt]{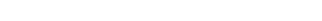}}
\fancyfoot[CO]{\vspace{-7.1pt}\hspace{11.9cm}\includegraphics{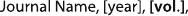}}
\fancyfoot[CE]{\vspace{-7.2pt}\hspace{-13.2cm}\includegraphics{head_foot/RF}}
\fancyfoot[RO]{\footnotesize{\sffamily{1--\pageref{LastPage} ~\textbar  \hspace{2pt}\thepage}}}
\fancyfoot[LE]{\footnotesize{\sffamily{\thepage~\textbar\hspace{4.65cm} 1--\pageref{LastPage}}}}
\fancyhead{}
\renewcommand{\headrulewidth}{0pt} 
\renewcommand{\footrulewidth}{0pt}
\setlength{\arrayrulewidth}{1pt}
\setlength{\columnsep}{6.5mm}
\setlength\bibsep{1pt}
%%%END OF FOOTER%%%

%%%FIGURE SETUP - please do not change any commands within this section%%%
\makeatletter 
\newlength{\figrulesep} 
\setlength{\figrulesep}{0.5\textfloatsep} 

\newcommand{\topfigrule}{\vspace*{-1pt}% 
\noindent{\color{cream}\rule[-\figrulesep]{\columnwidth}{1.5pt}} }

\newcommand{\botfigrule}{\vspace*{-2pt}% 
\noindent{\color{cream}\rule[\figrulesep]{\columnwidth}{1.5pt}} }

\newcommand{\dblfigrule}{\vspace*{-1pt}% 
\noindent{\color{cream}\rule[-\figrulesep]{\textwidth}{1.5pt}} }

\makeatother
%%%END OF FIGURE SETUP%%%

%%%TITLE, AUTHORS AND ABSTRACT%%%
\twocolumn[
  \begin{@twocolumnfalse}
{\hfill\raisebox{0pt}[0pt][0pt]{}\\[1ex]
}\par
\vspace{1em}
\sffamily
\begin{tabular}{m{4.5cm} p{13.5cm} }

 & \noindent\LARGE{\textbf{Molecular Mobility of Extraterrestrial Ices: Surface Diffusion in Astrochemistry and Planetary Science}} \\%Article title goes here instead of the text "This is the title"
\vspace{0.3cm} & \vspace{0.3cm} \\

 & \noindent\large{
 N.F.W. Ligterink,\textit{$^{a}$}  
 C. Walsh,\textit{$^{b}$} 
 H.M. Cuppen,\textit{$^{c}$}  
 M.N. Drozdovskaya,\textit{$^{d}$}  
 A. Ahmad,\textit{$^{b}$}  
 D.M. Benoit,\textit{$^{e}$}  
 J.T. Carder,\textit{$^{f}$}  
 A. Das,\textit{$^{g,h}$}  
 J. K. Díaz-Berríos,\textit{$^{b}$}  
 F. Dulieu,\textit{$^{i}$}  
 J. Heyl,\textit{$^{j}$}  
 A.P. Jardine,\textit{$^{k}$}  
 T. Lamberts,\textit{$^{l,m}$}  
 N.M. Mikkelsen,\textit{$^{n}$}  
 M. Tsuge,\textit{$^{o}$}  
 } \\%Author names go here instead of "Full name", etc.
\\

\includegraphics{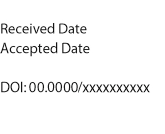} & \noindent\normalsize{Molecules are ubiquitous in space. They are necessary component in the creation of habitable planetary systems and can provide the basic building blocks of life. Solid-state processes are pivotal in the formation of molecules in space and surface diffusion in particular is a key driver of chemistry in extraterrestrial environments, such as the massive clouds in which stars and planets are formed and the icy objects within our Solar System. However, for many atoms and molecules quantitative theoretical and experimental information on diffusion, such as activation barriers, are lacking. This hinders us in unravelling chemical processes in space and determining how the chemical ingredients of planets and life are formed.
In this article, an astrochemical perspective on diffusion is provided. Described are the relevant adsorbate-surface systems, the methods to model their chemical processes, and the computational and laboratory techniques to determine diffusion parameters, including the latest developments in the field. While much progress has been made, many astrochemically relevant systems remain unexplored.
The complexity of ice surfaces, their temperature-dependent restructuring, and effects at low temperatures create unique challenges that demand innovative experimental approaches and theoretical frameworks. This intersection of astrochemistry and surface science offers fertile ground for physical chemists to apply their expertise. We invite the physical chemistry community to explore these systems, where precise diffusion parameters would dramatically advance our understanding of molecular evolution in space—from interstellar clouds to planetary surfaces—with implications on our understanding on the origins of life and planetary habitability.
} 

\end{tabular}

 \end{@twocolumnfalse} \vspace{0.6cm}
]

%%%END OF TITLE, AUTHORS AND ABSTRACT%%%

%%%FONT SETUP - please do not change any commands within this section
\renewcommand*\rmdefault{bch}\normalfont\upshape
\rmfamily
\section*{}
\vspace{-1cm}

%%%FOOTNOTES%%%
\footnotetext{\textit{$^{a}$~Faculty of Aerospace Engineering, Delft University of Technology, Delft, The Netherlands \\ E-mail:~niels.ligterink@tudelft.nl}}
\footnotetext{\textit{$^{b}$~School of Physics and Astronomy, Sir William Henry Bragg Building, University of Leeds, Leeds, LS2 9JT, UK \\ E-mail:~c.walsh1@leeds.ac.uk}}
\footnotetext{\textit{$^{c}$~Institute for Molecules and Materials, Radboud University, Nijmegen, The Netherlands \\ E-mail:~h.cuppen@science.ru.nl}}
\footnotetext{\textit{$^{d}$~Physikalisch-Meteorologisches Observatorium Davos und Weltstrahlungszentrum (PMOD/WRC), Dorfstrasse 33, 7260, Davos Dorf, Switzerland \\ E-mail:~maria.drozdovskaya.space@gmail.com }}
\footnotetext{\textit{$^{e}$~E.A.\ Milne Centre for Astrophysics, University of Hull, Hull, HU6 7RX, UK}}
\footnotetext{\textit{$^{f}$~Department of Chemistry, University of Virginia, McCormick Road, Charlottesville, VA 22904, USA}}
\footnotetext{\textit{$^{g}$~Institute of Astronomy Space and Earth Sciences, P 177, CIT Road, Scheme 7m, Kolkata 700054, West Bengal, India}}
\footnotetext{\textit{$^{h}$~Max-Planck-Institut für extraterrestrische Physik, Gießenbachstrasse 1, D-85748 Garching bei München, Germany}}
\footnotetext{\textit{$^{i}$~CY Cergy Paris Université, Observatoire de Paris, Université PSL, Sorbonne Université, Université Paris Cité, CNRS, LIRA, F-95000, Cergy, France}}
\footnotetext{\textit{$^{j}$~Department of Physics and Astronomy, University College London, Gower Street, WC1E 6BT, London, UK}}
\footnotetext{\textit{$^{k}$~The Cavendish Laboratory, University of Cambridge, JJ Thomson Avenue, Cambridge, CB3 0US, United Kingdom}}
\footnotetext{\textit{$^{l}$~Leiden Institute of Chemistry, Leiden University, Leiden 2300 RA, The Netherlands}}
\footnotetext{\textit{$^{m}$~Leiden Observatory, Leiden University, P.O. Box 9513, 2300 RA Leiden, The Netherlands}}
\footnotetext{\textit{$^{n}$~Center for Interstellar Catalysis, Department of Physics and Astronomy, Aarhus University, 8000 Aarhus C, Denmark; }}
\footnotetext{\textit{$^{o}$~Institute of Low Temperature Science, Hokkaido University, Sapporo, Japan}}

%%%END OF FOOTNOTES%%%

%%%MAIN TEXT%%%%

\section{Introduction}
\label{intro}

Astrochemistry studies the chemical makeup and molecular processes that occur throughout the Universe \cite{herbst2009complex}. Among these processes, surface diffusion—the movement of atoms and molecules on icy or dusty surfaces in space—emerges as a fundamental driver of chemical complexity. Understanding this phenomenon bridges molecular astrophysics, planetary science, and physical chemistry in an interdisciplinary area of research.

The chemical complexity we observe in space environments, from interstellar clouds to planetary surfaces, depends critically on solid-state reactions \cite{watson1972molecule, allen1977molecular,tielens1982model}. Unlike Earth-based chemistry, space chemistry faces unique constraints: gas and plasma phases are generally too dilute, with the synthesis of small, simple molecules like CO taking of the order of 10$^{5}$ years in interstellar clouds, and liquid phases are rare. On solid surfaces, however, higher density material is found and reactants can diffuse, meet, interact, and form larger molecules. Thus, surface diffusion plays a decisive role in determining reaction rates and ultimately shapes the molecular inventory of the Universe \cite{hase92, vandishoeck1998chemical}.

The historical development of this field began in the late 1940s when researchers first proposed that interstellar dust grains could serve as catalytic surfaces for molecule formation \cite{vandehulst1949solid}. This concept gained traction in the 1970s and 1980s as laboratory studies \cite{hagen1979interstellar,SandfordAllamandola1988} and theoretical models \cite{watson1972molecule,tielens1982model} began to illuminate the importance of surface processes. By the 1990s, sophisticated rate equation approaches \cite{hase92} emerged to model these complex surface-mediated reactions.

Direct measurement of diffusion in extraterrestrial environments remains impossible, requiring us to infer its importance through astronomical observations and laboratory analogues. Our understanding of diffusion has evolved through analysis of molecular abundances observed in diverse cosmic environments—from cold molecular clouds to planetary surfaces \cite{vandishoeck1998chemical,mcclure2023ice,clark2012observed}. For example, the ubiquitous water ice (\ce{H2O}) found throughout the Universe presents a compelling case for formation mediated by surface diffusion \cite{vandishoeck2013interstellar}.

\begin{figure}
\centering
\includegraphics[width=0.45\textwidth]{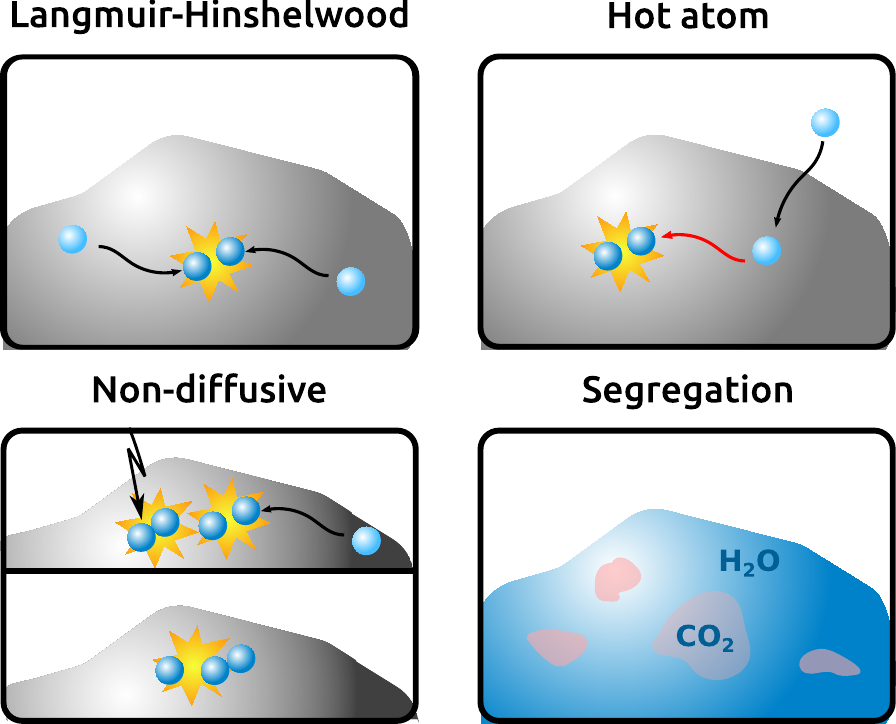}
\caption{Various examples of surface diffusion and related processes taking place in condensed phases in space. Each panel illustrates a distinct diffusion mechanism: Langmuir-Hinshelwood (top left) shows reactants diffusing before meeting; hot atom diffusion (top right) depicts energetic species moving rapidly across surfaces; non-diffusive reactions (bottom left) occur only requiring serendipitous proximity; and segregation (bottom right) shows separation of different molecular species in mixed ices. Image adapted from \citet{cuppen2024laboratory}.}
\label{fig:example2col}
\end{figure}

Water's emergence can be traced to the early stages of star and planet formation, beginning in molecular clouds—regions with relatively high density ($\gtrsim 10^{4}$ particles cm$^{-3}$) shielded from radiation. In these cold environments (approximately 10 K), gas-phase chemistry alone cannot explain the observed abundance of water ice. Instead, early gas-grain astrochemical models revealed that thermally activated diffusion of hydrogen and oxygen atoms on dust grain surfaces provides an efficient pathway for water formation \cite{tielens1982model} through the Langmuir-Hinshelwood mechanism (Figure \ref{fig:example2col}, top left panel).

Although molecular clouds consist predominantly of gas (with a typical gas-to-dust mass ratio of approximately 100), dust grains provide essential substrates for chemistry under these extreme conditions. Initially, diffusion of oxygen and hydrogen occurs on bare grains composed of silicates and carbonaceous material, possibly covered by hydrogenated amorphous carbon \cite{Jones:2013}. As ice accumulates, amorphous solid water surfaces become the dominant substrate for continued chemical reactions \cite{noble2024detection}. Freezeout of CO onto the icy dust grains on long timescales ($\sim$0.1 Myr) can be hydrogenated via hydrogen diffusion leading to the formation of \ce{CH3OH} in the ice phase\cite{linnartz2015atom}. \ce{CH3OH} provides a feedstock of key radicals (e.g., \ce{CH3}, \ce{CH3O}) which can go on to produce greater chemical complexity\cite{garrod2022formation}.  

These more complex molecules appear in the warmer environments ($\gtrsim$100~K) surrounding young stars. These regions, known as "hot corinos" (for low-mass protostars) and "hot cores" (for high-mass protostars), contain abundant complex organic molecules, including \ce{CH3OCH3} (dimethyl ether) and \ce{NH2CHO} (formamide) \cite{cazaux2003hot, bisschop2007testing, jorgensen2020astrochemistry, ligterink2020family}. These observations support a sequence where ice-coated dust grains migrate toward the forming star and experience heating. This warming allows previously immobile radicals within the ice to diffuse and recombine, forming larger molecules before eventually sublimating into the gas phase \cite{garrod2022formation}. The diffusion of radicals such as \ce{-CH3} (methyl), \ce{-NH2} (amine), and \ce{-CHO} (formyl) leading to functional groups to be added to molecules thus plays a crucial role in building the molecular complexity observed around young stars \cite{charnley1992molecular, garrod2022formation}.

\begin{figure}
\centering
\includegraphics[width=0.45\textwidth]{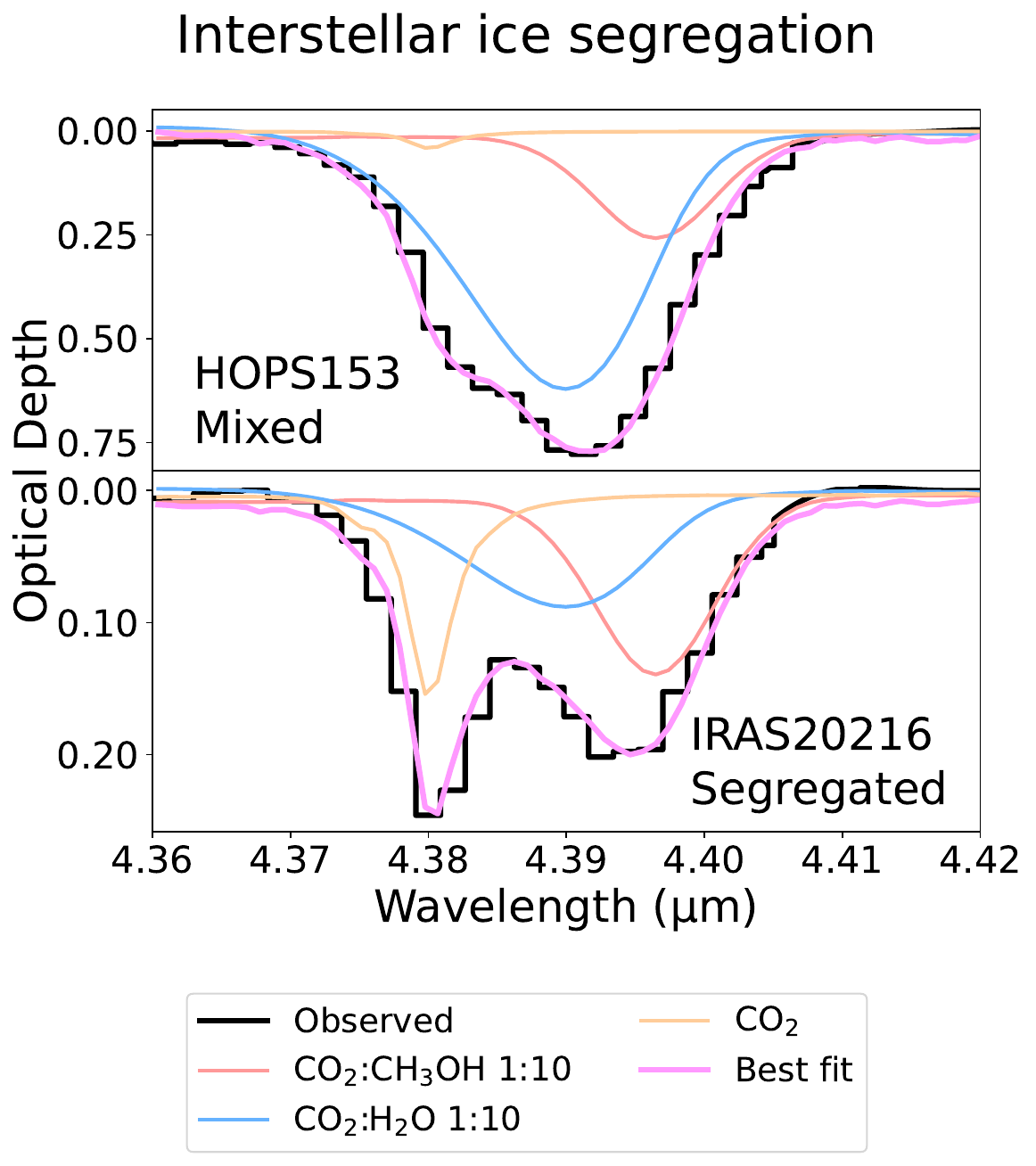}
\caption{The $^{13}$CO$_{2}$ asymmetric stretch vibrational mode observed in interstellar ice toward two different sources: HOPS153 (mixed/unsegregated ice) and IRAS20216 (segregated ice). The distinct spectral profiles reveal that segregation—a process driven by molecular diffusion—serves as a clear indicator of ice warming. Different molecular components (colored lines) contribute to the overall observed spectrum (black line). Data adapted from \citet{brunken2024jwst}.}
\label{fig:jwst_segregation}
\end{figure}

We can probe the solid phase of interstellar volatiles by analysing absorption spectra along lines of sight toward background stars \cite{boogert2015observations}. These observations reveal evidence of diffusion through characteristic spectral features. For example, infrared bands of solid \ce{CO2} display complex profiles (Figure \ref{fig:jwst_segregation}) that require fitting with multiple components, ranging from pure crystalline \ce{CO2} ice to intimately mixed \ce{CO2}:\ce{H2O} ices \cite{boogert2000iso, pontoppidan2008c2d, brunken2024jwst}. Such profiles indicate that ice undergoes segregation—separation of different molecular components—as it warms \cite{ehrenfreund1999laboratory}, a process that fundamentally depends on molecular diffusion (Figure \ref{fig:example2col}, bottom right panel).

Within our Solar System, diffusing ices leave observable signatures across various planetary bodies. On the Moon and Mercury, permanently shadowed regions act as cold traps where gases accumulate by freezing. These ice reservoirs persist thanks to surface diffusion on meter-scales into the regolith—the loose surface material overlying solid bedrock \cite{rubanenko2019thick}. On comets, both \ce{H2O} and \ce{CO2} ices display dynamic behaviour corresponding to diurnal and seasonal cycles \cite{DeSanctis2015, Filacchione2016, Filacchione2020}. During perihelion (closest approach to the Sun), cometary nuclei experience heating that enhances microscopic ice diffusion in subsurface layers \cite{SandfordAllamandola1988, willis2024ice}, contributing to the observed outgassing and activity.

From a fundamental chemical physics perspective, astrochemical systems offer the opportunity to investigate the mechanisms and implications of surface diffusion of relatively simple species on icy surfaces that are accessible in a laboratory setting. Despite its fundamental importance, accurate quantification of diffusion parameters—particularly activation barriers—remains one of the most significant challenges in astrochemistry \cite{cuppen2017grain,cuppen2024laboratory}. Current astrochemical models often rely on poorly constrained estimates that may vary by orders of magnitude, introducing substantial uncertainties in model predictions. Addressing this knowledge gap requires combined efforts from experimental physical chemistry, computational modelling, and astronomical observations.

This perspective paper addresses this critical research question: how can we better determine diffusion parameters for astrochemically relevant species and surfaces to improve our understanding of chemical evolution in space? We begin in Sect. \ref{sec:macscale} with an overview of surface diffusion treatment in astrochemical models, establishing key processes and terminology. We then explore the challenges in determining diffusion parameters from both experimental (Sect. \ref{sec:lab}) and computational perspectives (Sect. \ref{sec:molscale}). Finally, we conclude with a future outlook and summary of our perspective in Sects. \ref{sec:future} and \ref{sec:summary}, highlighting promising avenues for physical chemists to contribute to this expanding interdisciplinary field.

\section{Implementation of diffusion in astrochemical models}
\label{sec:macscale}

Astrochemical models serve as crucial bridges between microscopic chemical processes and macroscopic astronomical observations, allowing us to interpret observed molecular abundances and predict chemical evolution in space environments. These models need to cover time scales as long as 1$\times$10$^{5}$ to 1$\times$10$^{6}$ years, describing the chemical evolution of roughly 1,000 chemical species. The mean-field rate-equation model\citep{pick77,dhen85,hase92,shal94} provides the foundational framework for studying surface chemistry in interstellar environments. This approach employs ordinary differential equations to track molecular concentration changes over time, with the key assumptions of well-mixed reactants and continuous variables. At the  heart of the construction of these models lies a fundamental challenge: accurately representing how molecules move and interact on icy surfaces in the extreme cold of interstellar space.

The chemical evolution on grain surfaces typically follows a sequence of interconnected processes. Initially, gas-phase species accrete onto the grain surface. Once adsorbed, these species can diffuse across the surface, potentially encountering other reactants. When reactants meet, they may undergo chemical reactions to form new species. Whether specific reaction sites play a prominent role in forming the product is currently not clear for astrochemically relevant surfaces. Finally, surface species can desorb back into the gas phase through various mechanisms. These four processes—accretion, diffusion, reaction, and desorption—collectively determine the chemical composition of interstellar ices and, by extension, the gas-phase chemistry of molecular clouds.

When accretion exceeds the available binding sites on a grain, new ice layers form, causing the grain mantle to evolve from a single monolayer to a multilayer structure. This growth process has been modelled in increasingly sophisticated ways. In the two-phase model proposed by \citet{hase92}, all adsorbed materials are considered chemically reactive, with no distinction between species in the bulk ice or on its surface. In contrast, the more advanced three-phase model\citep{hase93b,furu15,ruau16} distinguishes between the chemically active surface layer and the less reactive mantle beneath. This distinction is crucial for realistic modelling, as reactivity decreases significantly in the ice bulk. Some modern models further refine this picture by treating several uppermost layers as active, rather than just the surface monolayer. This approach effectively incorporates surface roughness that increases the reactive surface area, providing a more realistic representation of actual interstellar ice surfaces, which observation and laboratory studies show are not perfectly smooth \citep{chen24}.

Once a species adsorbs onto a surface, it can either desorb (return to the gas phase), diffuse, or react. In the Langmuir-Hinshelwood mechanism, which dominates at low temperatures, reaction rates depend on the meeting frequency of reactants through their diffusion across the surface. This thermal hopping rate of a species across the surface is in rate equation models given by:

\begin{equation}
R_\mathrm{hop}={\nu_0} \exp\left(-E_\mathrm{diff}/k_\mathrm{B}T_\mathrm{dust}\right),
\label{eqn:hop}
\end{equation}

\noindent where $k_\mathrm{B}$ is Boltzmann's constant, $T_\mathrm{dust}$ is the dust grain temperature, and $\nu_0$ is the pre-exponential factor of the adsorbed species in its binding site. This factor can be associated with the frustrated translation vibrational mode frequency in the transition state theory (TST) limit, whcih represents how often the molecule "attempts" to overcome the potential energy barrier between adjacent sites. Several assumptions are made in this equation, such as TST and a single binding energy for each site. 

In astrochemical models, the diffusion rate is determined by dividing the hopping rate by the total number of binding sites on the grain. This astrochemistry-specific relationship is expressed as $R_\mathrm{diff} = R_\mathrm{hop}N_\mathrm{s}^{-1}$, where $N_\mathrm{s}$ is the number of binding sites per grain (typically $\sim 10^{6}$ for a grain with a radius of 0.1 $\mu$m). For lighter species, particularly H, D and \ce{H2}, quantum tunnelling can become more efficient than thermal hopping at very low temperatures, below its "crossover temperature"\citep{senevirathne2017hydrogen}, although its impact on long-range diffusion is limited. It is often assumed - for simplicity - that the barrier shape for quantum tunnelling is rectangular \citep[][]{hase92}, although some astrochemical modelling codes \citep[][]{Taquet2013} assume a more realistic Eckart potential \citep[][]{Eckart1930}.

Surface diffusion directly impacts reaction rates in astrochemical models. The reaction rate between species $i$ and $j$ on the surface depends on their meeting frequency in a common binding site, expressed as:

\begin{equation}
R_{ij}=\kappa_{ij}(R_{\mathrm{diff},i}+R_{\mathrm{diff},j})N_i N_j n_\mathrm{d}.
\end{equation}

\noindent where $N_i$ and $N_j$ are the numbers of reactants $i$ and $j$, respectively, on the surface of an average grain, $\kappa_{ij}$ is the probability of reaction upon encounter, and $n_\mathrm{d}$ is the number density of dust grains. For reactions without an activation barrier, the probability of reaction upon encounter ($\kappa_{ij}$) is typically assumed to be unity. However, for reactions with an activation barrier $E_\mathrm{a}$, this probability is given by the Arrhenius equation:

\begin{equation}
\kappa_{ij} = \exp\left(-E_\mathrm{a}/kT_\mathrm{dust}\right).  
\end{equation}

For reactions involving light species like hydrogen, quantum mechanical tunnelling through the barrier becomes important, and under the assumption of a rectangular barrier shape, the reaction probability changes to:

\begin{equation}
    \kappa_{ij}=\exp[-2(a/\amshbar)\sqrt{2\mu E_a}],
\end{equation}
\noindent where $\mu$ is the reduced mass and $a$ is the barrier width, most often chosen to be 1 \AA.

Given the lack of laboratory measurements, educated estimates are often required for many parameters mentioned above. A large gas-grain chemical network may include over 10,000 reactions involving approximately 1,000 species, with roughly one-third in the ice phase. For each ice species, modellers need values for the desorption barrier, diffusion barrier, pre-exponential factor, and barrier shape/width, and often employ educated estimates. One common approach assumes the diffusion barrier ($E_\mathrm{diff}$) to be a fixed fraction of the desorption barrier. Literature values for this ratio typically range from 0.3 (fast surface diffusion) to 0.7 (slower bulk diffusion) \citep{hase92,karssemeijer2014diffusion,garrod2022formation}. However, this broad parametrization is an approximation, as we know that not all species share a common ratio for diffusion and desorption barriers\citep{hedgeland2009measurement,furuya2022diffusion}. Further research into the generality of this ratio is needed to use it with more confidence.

A second significant challenge involves the pre-exponential factor ($\nu_0$) used in the rates for both desorption and diffusion. In laboratory measurements, the pre-exponential factor and diffusion or desorption energy are determined simultaneously by fitting experimental data. Consequently, the derived barrier for desorption (or binding energy) should ideally be used together with its associated pre-exponential factor. For desorption, most models use the harmonic oscillator expression to estimate pre-exponential factors, but the difference between calculated versus laboratory-derived values can span orders of magnitude \citep{mini22,ligt23}. Furthermore, astrochemical models typically assume that the pre-exponential factor for diffusion equals that for desorption. While this assumption is unlikely to be correct, it is a pragmatic approach to overcome the lack of experimental data. In turn, this highlights the need for accurately determined pre-exponential factors for diffusion. 

Recent work by \citet{chen24} demonstrates how sensitive molecular abundances are to these parameters. Their study found that variations in the diffusion pre-exponential factor can change predicted abundances of certain species by orders of magnitude. They showed that complex organic molecules (COMs) like \ce{CH3OCH3} and \ce{NH2CHO} are particularly sensitive to diffusion parameters, with abundance variations exceeding a factor of 10 under different parameter choices. These results highlight why accurate diffusion parameters are essential for reliable astrochemical modelling, especially when comparing predictions to observations of COMs in star-forming regions.

Astrochemical models contain inherent uncertainties in their computation of diffusion barriers and rates, which propagate to uncertainties in grain-surface reaction rates and ultimately to predicted molecular abundances. These uncertainties directly impact our ability to interpret astronomical observations of molecular species in cold cloud environments, star-forming regions, and protoplanetary disks. The next section outlines experimental approaches used to measure diffusion parameters, providing crucial constraints for astrochemical models and helping to reduce these uncertainties.

\section{Experimental studies}
\label{sec:lab}

Diffusion under astrochemically relevant conditions has been studied in laboratories for various adsorbates (analytes) and surfaces, focusing on determining diffusion constants needed for astrochemical models. This section details astrochemical experiments, the simulation of space conditions and surfaces, and techniques for measuring diffusion coefficients of adsorbed species.

The space environments relevant to diffusion studies span a range of conditions that laboratories must carefully replicate. These environments are generally characterised by low pressure ($\leq$10$^{-6}$ mbar) and cold surface temperatures (from a few Kelvin up to the water freezing point). A typical experimental astrochemical setup comprises a vacuum chamber and cryogenically cooled surface, often utilising a closed-cycle liquid helium or liquid \ce{N2} cryostat to maintain the extremely low temperatures found in interstellar space. Pre-prepared surfaces like cleaved olivine ([Mg,Fe]\ce{SiO4}) \cite{minissale2014oxygen} or highly oriented pyrolytic graphite (HOPG) \cite{minissale2016direct} can be mounted on the cryostat to serve as substrates for ice formation. Alternatively, laser-ablation of carbon targets followed by gas-phase condensation can be used to create highly amorphous and porous carbon structures \cite{jager2008spectral} that more closely resemble the disordered carbonaceous materials found in interstellar environments. These substrates represent different types of interstellar dust grains, which serve as the nucleation sites for ice formation in molecular clouds.

Water ice surfaces, the most abundant solid-phase material in interstellar environments, can be created $in-situ$ in various forms with substantially different properties. Amorphous solid water (ASW) of varying porosity is formed when water vapour freezes on a pre-cooled cryostat surface\cite{jenniskens1995high,noble2024detection}. The precise structure and morphology of this ice significantly affects diffusion behaviour and it depends on several factors, such as the deposition method (whether background filling or directed deposition), deposition rate, substrate temperature during deposition, and deposition angle\cite{stevenson1999controlling}. Post-deposition processing, such as annealing, can further modify the ice properties, specifically to reduce the porosity or create crystalline ice. Advanced techniques like matrix sublimation \cite{kouchi2016matrix} provide additional control over ice structure. Similar preparation methods apply to other volatile molecules commonly found in interstellar ices, such as CO, \ce{CO2}, and various hydrocarbons. The ability to create well-characterised ice analogues with reproducible properties is crucial for obtaining meaningful diffusion measurements that can be applied to astrochemical models.

Once the surface is prepared, diffusion studies of an adsorbate can be started. Direct visualisation techniques provide the most straightforward evidence of molecular diffusion. \citet{kouchi2020direct} developed a powerful method using ultra-high-vacuum transmission electron microscopy (UHV-TEM), where diffusion activation energy is determined from aggregate growth rates on a substrate. This technique relies on the principle that as molecules diffuse across a surface, they eventually encounter others and form aggregates that can be directly observed. The UHV-TEM method applies primarily to stable molecules, as free radicals react too quickly to form observable aggregates. Because aggregates are visually counted from TEM images, adsorbates that form crystalline aggregates on an amorphous substrate, which give high-contrast images, are desirable. This approach has been applied to a number of adsorbates \cite{furuya2022diffusion,kouchi2021morphology} to determine their diffusion activation energies ($E_{\rm diff}$). For example, this method established that $E_{\rm diff}$ for CO molecules diffusing on an ASW surface is approximately 350 K (30 meV).

\begin{figure*}
\includegraphics[width=1\textwidth]{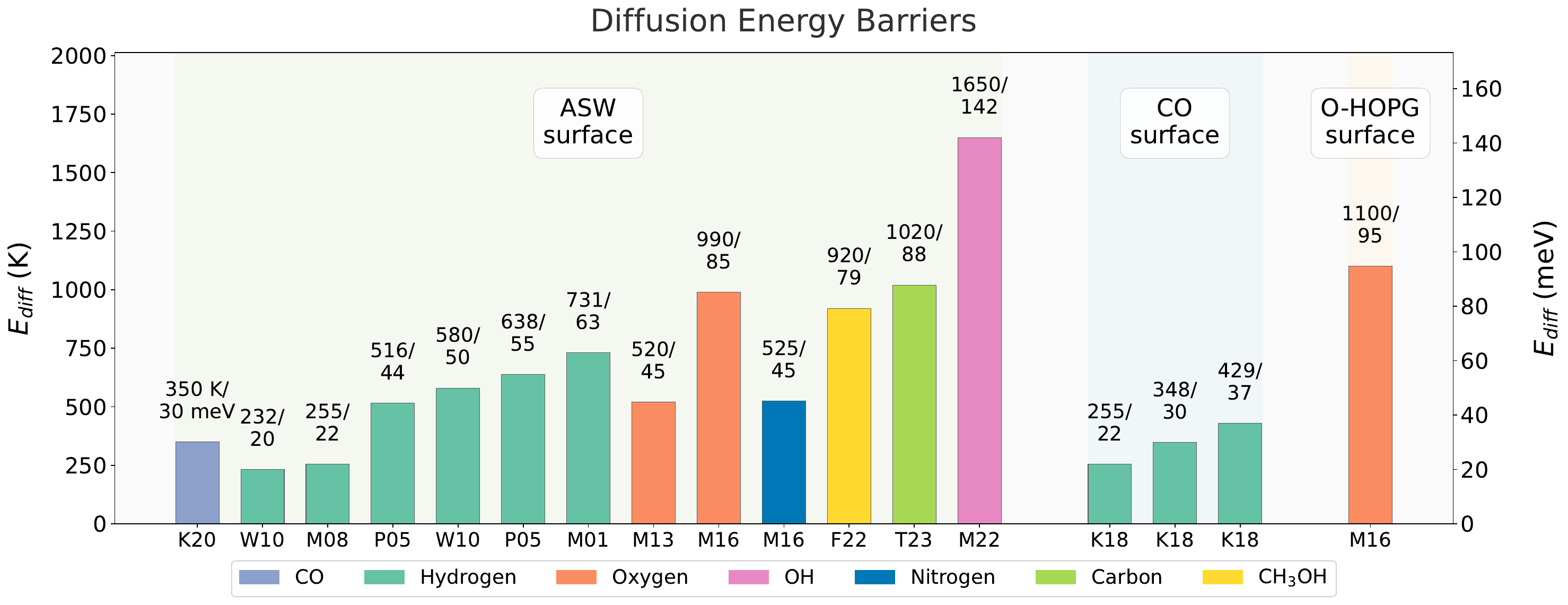}
\caption{Overview of experimentally determined diffusion energies for various atoms and molecules on different surfaces in K (left y-axis) and meV (right y-axis). The bar colour indicates the diffusing species, which are grouped in the categories ASW = Amorphous Solid Water, CO = carbon monoxide, and O-HOPG = oxygenated highly oriented pyrolytic graphite. We note that a relatively large amount of data is available for diffusion on water-ice, but data for carbonaceous, siliceous, and other ice surfaces is scarce, as the plot underlines. Various subdivision of the ice surfaces are made based on their porosity (non-porous, porous, compact), density (low, high), temperature, and depth of binding sites (deep, shallow), which drive the differences in diffusion barriers. Data is taken from M01\citep{manico2001laboratory} (ASW), P05\citep{perets2005molecular} (low and high density ASW), M08\citep{matar2008mobility} (porous ASW), W10\citep{watanabe2010direct} (shallow ASW), M13\citep{minissale2013quantum} (ASW), M16\citep{minissale2016direct} (compact ASW, O-HOPG), K18\citep{kimura2018measurements} (8, 12, 15K CO), K20\citep{kouchi2020direct} (porous ASW), F22\citep{furuya2022diffusion} (compact ASW), M22\citep{miyazaki2022direct} (ASW), T23\citep{tsuge2023surface} (non-porous ASW).
\label{fig:exp-diff}}
\end{figure*}

Figure \ref{fig:exp-diff} shows a selection of diffusion values determined with different experimental techniques for various adsorbates and surfaces. These measurements reveal significant variations in diffusion barriers depending on both the diffusing species and the substrate characteristics, highlighting the importance of species-specific and surface-specific parameters for astrochemical models.

While direct visualisation provides clear evidence of diffusion, many experimental setups utilise spectroscopic or mass spectrometry techniques to infer diffusion parameters. A common protocol employs isothermal infrared (IR) spectroscopy measurements on layered or mixed ices. In these experiments, a system consisting of a bulk component (commonly \ce{H2O} or \ce{CO2}) and a target analyte molecule arranged in layers or mixed is prepared. When the ice sample is heated to a specific temperature, the target species begins to diffuse through the bulk material. This movement alters the molecular interactions, which in turn affects the infrared absorption spectra in measurable ways. With these experiments 3D diffusion or surface diffusion through open pores is traced, rather than pure surface diffusion. 

Changes in IR signatures of the target or bulk species are monitored to track diffusion. Examples include depletion of the dangling OH bond of water \cite{mate2020diffusion} as diffusing molecules interact with these sites, variations in the longitudinal optical phonon mode of \ce{CO2} \cite{cooke2018co}, surface cluster formation \cite{he2017diffusion}, or IR band depletion due to desorption when diffusing molecules reach the ice surface \cite{mispelaer2013diffusion,karssemeijer2013dynamics,ghesquiere2015diffusion}. The temporal evolution of these spectral changes are fitted with diffusion models to extract the diffusion coefficients. This approach has gained widespread use due to its relative simplicity and the common availability of IR spectrometers in astrochemical laboratories.

Mass spectrometry-based techniques offer complementary capabilities for studying diffusion. Laser desorption methods \cite{livingston2002general} use laser pulses to "excavate" material from specific locations on the ice surface and transfer it to the gas phase for analysis by mass spectrometry. This approach can measure concentration gradients of target molecules diffusing through a bulk medium, providing another pathway to determine diffusion rates. An alternative approach deposits an adsorbate uniformly on a surface, uses a laser to clear a specific area, and then measures the signal as surrounding molecules diffuse into the cleared region. The mass spectrometry approach offers advantages for studying species that are difficult to detect with infrared spectroscopy, such as noble gases and other volatiles, e.g., \ce{N2}, and \ce{O2}. It also allows diffusion measurements on non-ice substrates that would be challenging to study with IR techniques alone. 

While the techniques described above work well for stable molecules, studying the diffusion of atomic and molecular radicals presents unique experimental challenges. Radicals are highly reactive species with unpaired electrons, making them difficult to prepare, control, and detect due to their short lifetime. Yet understanding radical diffusion is crucial because interstellar ice chemistry depends heavily on radical-radical and radical-molecule reactions. Atomic radicals such as H, O, and N are typically produced in beam lines by plasma or radio frequency dissociation of precursor molecules \cite{he2014atomic,minissale2013quantum}. For example, H atoms can be generated from H$_2$ dissociation, O atoms from O$_2$, and so forth. These radicals are then directed onto the prepared surface, where their diffusion behaviour must be studied before they react to form stable products.

Measuring the diffusion process of free radicals is challenging because of the difficulty of $in-situ$ detection of radicals on a substrate. One approach to studying radical diffusion involves an indirect measurement strategy. Radicals are deposited on a surface held at a specific temperature, given time to diffuse and react, and then the reaction products are measured. Generally, these measurements involve gas-phase detection of products using mass spectrometry, often combined with temperature-programmed desorption (TPD) techniques. By fitting experimental data to models that incorporate diffusion rates, reactant concentrations, and reaction kinetics, the diffusion parameters can be extracted.

This approach has been successfully applied to determine diffusion parameters for H atoms \cite{manico2001laboratory,perets2005molecular,matar2008mobility}, O atoms \cite{congiu2014efficient,minissale2014oxygen,he2014atomic,he2015new,minissale2016direct}, and N atoms \cite{minissale2016direct} on various surfaces. Multiple effects need to be considered in the multi-parameter fit routine, while some are poorly constrained. Moreover, TPD is not suitable to analyse irregular amorphous surfaces, where the activation energy cannot be represented by a single value but has a wide distribution, as the free fit parameter space becomes too large. The results show considerable variation, as illustrated by the range of hydrogen diffusion barriers presented in Figure \ref{fig:exp-diff}.

A more direct method for studying radical diffusion combines photostimulated desorption (PSD) with resonance-enhanced multiphoton ionisation (REMPI)\citep{watanabe2010direct}. Surface-adsorbed species are desorbed using a weak nanosecond laser pulse, then ionised using the REMPI method, and finally detected with a time-of-flight mass spectrometer. This approach offers advantages of high sensitivity, surface selectivity, and species-specific detection.

The PSD-REMPI technique has been successfully applied to study diffusion of H atoms \cite{hama2012mechanism,kuwahata2015signatures,kimura2018measurements}, C atoms \cite{tsuge2023surface}, and molecular OH radicals \cite{miyazaki2020photostimulated,tsuge2021behavior,miyazaki2022direct}. To study surface diffusion, the time variation of the surface number density is recorded; thus, only the surface diffusion of species that react away upon meeting can be studied by this method. Moreover, it would be difficult to study larger strongly bound species because PSD might not be able to desorb these species. Ideally we would have access to quantitative information at the atomistic level.

Despite significant progress in experimental techniques for measuring diffusion in astrochemically relevant systems, many challenges remain. Creating truly representative analogues of interstellar ices is difficult, particularly for mimicking the complex, multi-component ices that exist in space environments. Further development of methods that can access a wider range of molecules, surfaces, and conditions will be essential for building a comprehensive understanding of diffusion processes in astrochemical systems.

The high variability in measured diffusion parameters for the same species across different studies (as illustrated in Figure \ref{fig:exp-diff}) highlights the sensitivity of diffusion to precise surface conditions and measurement techniques. Future work will need to focus on reconciling these variations and establishing standardised approaches that produce consistent, reliable diffusion parameters for use in astrochemical models.

\section{Computational studies}
\label{sec:molscale}

Besides observations and experimental measurements, computational approaches also provide a useful insight into diffusion phenomena at the molecular scale. Moreover, computational chemistry models not only describe molecules or substrates that are hard to isolate or prepare in the laboratory, but also have the ability to do so at the molecular or atomistic level of detail. To achieve a high enough accuracy at a reasonable cost, often-used methods operate at a force-field \citep{ Batista:2001, karssemeijer2013dynamics, karssemeijer2014diffusion,pezzella2018molecular} or density-functional theory level \citep{enrique-romero2025interstellar}. Recently, machine-learned potentials operating at DFT quality are starting to become feasible \citep{zaverkin2022neural}. Alternatively, a potential can be fitted specifically for the system at hands based on high level (e.g., coupled cluster theory) single point energies \citep{asgeirsson2017long,senevirathne2017hydrogen,ferrari2023floating}. 

\begin{figure}
\includegraphics[width=0.5\textwidth]{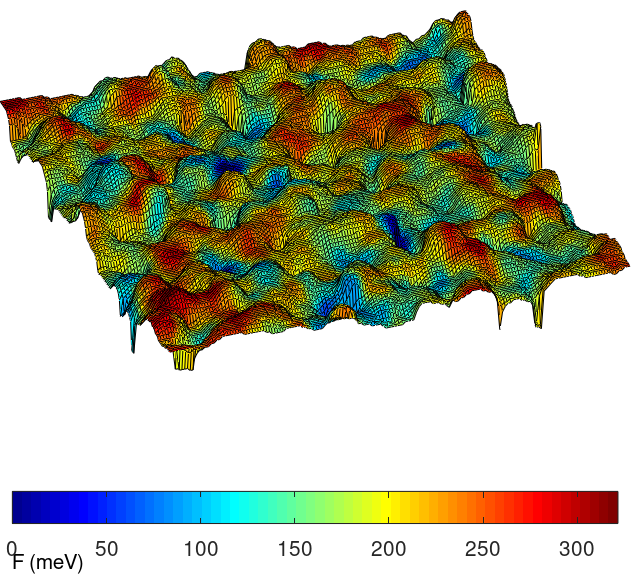}
\caption{Result from a metadynamics simulation of the diffusion of a \ce{CO2} molecule over an ASW surface. The surface mesh indicates the van-der-Waals surface of the ASW surface while the color coding represents the interaction free energy of the \ce{CO2} with the surface at 30~K. The most stable locations coincide with the deeper areas on the surface. Diffusion pathways between sites avoid surface protrusions.}
\label{fig:fes}
\end{figure}

Theoretically, the diffusion process can be depicted as a ``reaction path'' originating at one of the adsorption sites, going through an energetically less favourable region (transition state for diffusion) and ending at another site (or desorbing). The parallel between a diffusion process and a chemical reaction means that many computational tools used for mapping chemical reactions at a molecular level can also be used to explore diffusion. Nudged-elastic band has for instance been employed to obtain diffusion barriers of \ce{H2O} on a crystalline \ce{H2O} surface \citep{Batista:2001} and very recently for \ce{CH3} on ASW \citep{enrique-romero2025interstellar} or bulk diffusion of small molecules in ASW has been studied by molecular dynamics \citep{ghesquiere2015diffusion}.
One main hurdle is that, in a typical diffusion process on an amorphous solid, such as interstellar ice, the actual adsorption surface and/or the interaction mode is often ill-characterised. However, there is active research in employing machine learning approaches to obtain better surface models, also for amorphous systems (see for example \citet{ronne2024generative}). Amorphicity leads to very complex energy landscapes, see for example Fig.~\ref{fig:fes} which depicts the free-energy landscape for \ce{CO2} on ASW. The many different available minima and transition states can derail exploration techniques developed initially for gas-phase chemistry. 
Different sampling techniques such as metadynamics can help to map out this landscape and has been applied for diffusion by a number of groups \citep{zaverkin2022neural,Ward_2019,Pramanik2020}. Once a distribution of diffusion barriers is obtained the kinetic Monte Carlo method can be applied to arrive at a diffusion coefficient that describes the long-range diffusion of species over the surface, in addition to measurements of the site-to-site hopping that can keep the adsorbate within a local surface pocket \citep{zaverkin2022neural,Batista:2001,karssemeijer2013dynamics,karssemeijer2014diffusion}.

Long-range diffusion simulations \citep{asgeirsson2017long} and binding and diffusion barrier calculations \citep{senevirathne2017hydrogen} show that long-distance diffusion of H-atoms on an ASW surface is determined by the thermal rate from the deep wells. Since this is most typically a local endothermic process, the role of tunnelling is minimal. However, if the deep sites are blocked by other adsorbates, tunnelling becomes important for the diffusion between remaining shallow sites. This is consistent with experiments that show that H diffusion on ASW is highly coverage-dependent and much more efficient for high coverage \citep{kuwahata2015signatures}.

Kinetic Monte Carlo studies of CO and N diffusion also show that blocking of deep sites can increase diffusion \citep{zaverkin2022neural,karssemeijer2014interactions}. This is because diffusion from shallower sites have a lower barrier, and not because of a difference in tunnelling versus thermal behaviour. Indeed these heavy species are not expected to be affected by tunnelling for their mobility. Using a combination of molecular dynamics and Monte Carlo simulations, \citet{pezzella2018molecular} found that quantum tunnelling is not required to explain the experimental results \citep{congiu2014efficient} for mobility of adsorbed oxygen on ASW, but the increased mobility at low temperatures could be accounted for by including surface roughness and a distribution of binding sites.

Ultimately, regardless of the theoretical approach adopted for obtaining a diffusion barrier or rate, calculations rely on an accurate description of the molecule--surface interactions, or at least a description that captures enough detail to model the process. This, however, remains a very challenging task due to multiple minima (sites) on the surface (see Fig.~\ref{fig:fes}), difficulties in deciding how to model the actual solid surface studied (ASW has many different computational models, for example \cite{Germain_2020, Molpeceres_2020, Ferrero_2020, Woon_2023}), dealing with dispersion-dominated interaction \citep{ferrari2023floating,mikkelsen2025trends} and in choice of method used to describe the overall diffusion energetics \cite{Fredon_2018,Sameera_2020, Ferrero_2020, Bovolenta_2020, Molpeceres_2021,Duflot_2021, Clark_2024}. 

At present, the application of machine learning or new low-cost total energy methods are still quite rare and is set to expand the horizon of what size of system can be studied in reasonable computational times. There is also a huge research effort to develop fast high-accuracy methods (exemplified by the success of recent local approaches to highly-correlated methods, such as the domain-based local pair natural orbital-DLPNO-methods family) which will offer a better description of the potential energy surface at much lower cost. Those two types of advances alone would help exploring diffusion of larger molecules (COMs, for example) on more complex substrates (mixed amorphous ices). One emerging theme for those studies is the description of diffusion of radicals on astrochemically-relevant surfaces, since those are typically hard to create in the laboratory but also remain challenging to model theoretically. \\

\section{Future outlook}
\label{sec:future}

In the past decades, researchers have made significant advances in understanding diffusion processes under astrochemical conditions, yet several barriers hinder our ability to accurately model chemical evolution in interstellar environments. Laboratory measurements of atoms and radicals remain challenging and computational methods are hampered by lengthy calculation times. In general, accurate data on diffusion parameters remains scarce and measurements of a wider variety of adsorbates and surfaces remain necessary, necessitating new and innovative experimental and computational techniques. The following sections outline specific approaches that will drive this next wave of discovery.

\subsection{New tools}

One prospective method for studying radical diffusion in the laboratory is the \ce{Cs^+}-ion pickup method,\cite{tsuge2023review} in which surface adsorbates are picked up by a low-energy \ce{Cs^+}-ion beam and analyzed by a quadrupole mass spectrometer. This method has been demonstrated to be capable of sensitively detect a trace amount of free radicals or COMs on ASW;\cite{ishibashi2021pickup} at a detection limit is as low as $10^{-4}$ monolayer coverage ($1 \times 10^{11}$ molecules cm$^{-2}$). With this method, one can study surface diffusion by tracing the time variation of the number density of free radicals similar to PSD-REMPI. One advantage of this method over PSD-REMPI is its capability of detecting larger species, however without the ability to distinguish between structural isomers due to its mass spectrometric detection. Thus, \ce{Cs^+}-ion pickup and PSD-REMPI methods could be complementary to each other in studying surface diffusion of free radicals.

Helium microscopy (SHeM) and helium-3 spin-echo ($^{3}$HeSE) techniques offer complementary approaches for non-destructive surface analysis using neutral atom probes. SHeM generates images and performs diffractive/TPD spot-profile analysis, using the helium interaction with the outermost electronic structure of the samples. The low beam energy makes it ideal for delicate and insulating materials, eliminating the possibility of electron stimulated damage or dissociation or species such as water, and providing high sensitivity to small species, including H atoms\cite{barr2016unlocking,vonjeinsen20232d,zhao2025multi}. $^{3}$HeSE measures correlations in nanoscale surface motion, which makes it ideal for studying rates and mechanisms of surface diffusion\cite{tamtogl2021motion,kyrkjebo20233he}. The method works by using the nuclear spins in the beam of $^{3}$He atoms to perform a temporal interferometric measurement on the surface; essentially each helium atom’s wavefunction is split and separated in time, scattered from the surface, then recombined.  The final spin-polarisation corresponds to the change in surface correlation, thus revealing surface dynamics on picosecond timescales and nanometre length scales. These techniques\footnote{CORDE Atom scattering Facility, Cambridge, https://corde.phy.cam.ac.uk.} have revealed diffusion coefficients and activation energies for water molecules on graphene\cite{kyrkjebo20233he}, have provided insights into kinetic barriers to ice nucleation\cite{tamtogl2021motion}, and have studied mechanisms of H diffusion on surfaces\cite{jardine2010determination}, making them promising options to investigate diffusion in astrochemically-relevant systems.

In experiments, atomic radicals such as H, N, and O are typically generated in a plasma from \ce{H2}, \ce{O2}, and \ce{N2}, respectively. Unfortunately, we cannot apply this method to the production of important atomic radicals like S and P. The primary choice for the production of these species would be the in-situ photolysis of \ce{H2S} and \ce{PH3} precursors. The production of other important molecular radicals such as \ce{NH2}, NH, \ce{CH3}, and \ce{CH2} is also challenging because a photolysis method will yield mixtures; e.g., \ce{NH2}, \ce{NH}, and N in the photolysis of \ce{NH3}. To selectively produce these molecular radicals, there are two possible methods that have not been exploited in solid-phase astrophysical studies. One is the pyrolysis (thermal cracking) method,\cite{kohn1992pyrolysis} where appropriate precursor molecules are dissociated in a hot nozzle and the produced radicals are deposited onto a substrate. For example, \ce{CH3} can be produced by the pyrolysis of \ce{CH3I} because the C-I bond is weakest in \ce{CH3I}. The other method is the neutralization of ions. Because mass selection can be achieved for ions, one can isolate a beam of single radical species. To apply this method, a deceleration system for the generated radical beam is required. Finally, the radical production method used by Ishibashi et al.\cite{ishibashi2021pickup} is noted here. In their protocol, an ASW sample was photolyzed by a conventional deuterium lamp to produce OH radicals on the surface and a precursor molecule (XH) was subsequently deposited; radical species "X" will be produced via the reaction \ce{OH + XH -> H2O + X}, if the reaction is exothermic.     

\subsection{Machine Learning}

Gaps in known diffusion parameters can be alleviated with the use of machine learning (ML). ML has quickly become a powerful tool in astrochemistry, primarily as a computationally cheap alternative to quantum chemistry calculations, as mentioned above. The ability of ML to interpret and find patterns in large amounts of data makes it possible to incorporate experimental data when trying to predict parameters of new systems. For example, \citet{villadsen2022predicting} use binding energies obtained using temperature-programmed desorption as data to develop a model that predicts binding energies of new molecules of astrochemical relevance. Unfortunately, we currently lack enough diffusion barrier data to attempt a similar approach for diffusion barriers.

There is often a danger of these machine learning algorithms acting as ``black boxes'' in that the relationship between the input parameters and output parameters is not necessarily interpretable to practitioners. The various parameters (features) may also have range-dependent effects on the binding energy values. It is for precisely these scenarios that we may require an understanding of the feature importance. SHapley Additive exPlanations (SHAP), a machine learning methodology based on Shapley values from game theory, allow for the quantification of the various features by treating the individual features as providing an additive contribution to the value of the final output \cite{Shapley, lundberg2017unified, lundberg2018consistent, molnar2022}.

Further to all of this, there may still be a need to obtain accurate functional forms for parameters such as the pre-factors and the diffusion/desorption energies. For these scenarios, symbolic regression approaches can be utilised to relate quantities of interest to the pre-factors and diffusion/desorption energies. Symbolic regression is a regression method that explores function spaces to identify regression models that fit the data well \cite{Angelis2023}. The purpose of obtaining algebraic expressions is twofold. In the first instance, these expressions can aid in interpretability and can be reconciled with physics/chemistry. Additionally, these expressions can be inserted in to astrochemical codes to improve modelling efforts. 

\section{Conclusion}
\label{sec:summary}

Surface diffusion plays an important role in shaping the chemical inventory in the interstellar medium and solar system bodies. Recent experimental and computational studies have shown success in constraining diffusion parameters under astrochemically-relevant conditions. However, significant knowledge gaps remain, specifically for atoms, radicals, and many of space-relevant surfaces. It remains unclear whether diffusion barrier distributions needs to be considered for the mobility of species or surface chemistry can be characterized only using a single value, be it the average, upper, or lower bound of the distribution. In general, more accurate data on diffusion parameters is required to inform the expansive chemical networks used in astrochemical models. This drives the need for novel experimental approaches and (machine learning) enhanced computational methods. We invite and encourage the physical chemistry and chemical physics community to explore these systems, where precise diffusion parameters would dramatically advance our understanding of molecular evolution in space from interstellar clouds to planetary surfaces. 

\section*{Acknowledgements}
This perspectives article arose in thanks to an interdisciplinary workshop on surface diffusion held at the University of Leeds in April 2024 that was supported by funding from UK Research and Innovation (grant number MR/T040726/1). The authors thank Nashanty Brunken for providing JWST spectra of segregated and unsegregated interstellar CO$_{2}$ ice. N.F.W.L. acknowledges support from the Swiss National Science Foundation (SNSF)
Ambizione grant 193453. C.W.~acknowledges financial support from the Science and Technology Facilities Council and UK Research and Innovation (grant numbers ST/X001016/1 and MR/T040726/1). J.~K.~D.-B. acknowledges support from the Science and Technology Facilities Council via a doctoral training grant (grant number ST/Y509711/1). N.M.M. acknowledges funding from the Danish National Research Foundation through the Center of Excellence "InterCat" (Grant agreement no.: DNRF150). M.T. acknowledges funding from JSPS KAKENHI Grant Number JP24K00686. 

\balance

%%%REFERENCES%%%
\bibliography{rsc} 
\bibliographystyle{rsc} 

\end{document}